\documentstyle[12pt]{article}
\newcommand{\be}{\begin{eqnarray}}
\newcommand{\ee}{\end{eqnarray}}
\begin{document}
\thispagestyle{empty}
\begin{center}

\vspace*{3cm}

{\large{\bf Non-Perturbative Regge
Exchange in Meson-Meson Scattering:\\
An Analysis Based on the Stochastic Vacuum Model\\}}

 \vspace*{1.5cm}

{\large A.~I.~Karanikas and C.~N.~Ktorides}\\
\smallskip
{\it University of Athens, Physics Department\\
Nuclear \& Particle Physics Section\\
Panepistimiopolis, Ilisia GR 15771, Athens, Greece}\\
\vspace*{2.5cm}

\end{center}
\begin{abstract}

Employing the Worldline casting of the Dosch-Simonov Stochastic
Vacuum Model (SVM) for QCD, a simulated meson-meson scattering
problem is studied in the Regge kinematical regime. The process is
modelled in terms of the $`$helicoidal' Wilson contour first
introduced by Janic and Peschanski in a related study based on an
AdS/CFT-type approach. Using lattice supported estimations for the
behavior of a two-point, field strength correlation function, as
defined in the framework of the SVM, the reggeon slope and
intercept are calculated in a semiclassical approximation. The
resulting values are in good agreement with accepted
phenomenological ones. Going beyond this approximation, the
contribution resulting from boundary fluctuations of the Wilson
loop contour is also estimated.

\end{abstract}

\newpage

{\bf 1. Introductory Remarks}

In addition to confinement, which constitutes a profoundly
non-perturbative, problem and whose solution is of quintessential
importance, for fully establishing QCD as a fundamental theory for
the strong interaction, there do exist specific {\it dynamical}
processes, whose theoretical confrontation also calls
 for non-perturbative methods of analysis.
 One such situation arises in connection with the theoretical description of
high energy scattering amplitudes for which the soft sector of the
theory is involved. From the experimental point of view, one such
case arises in connection with Regge kinematics, entering directly
the theoretical description of, among others, diffractive and
low-x physics processes. In this paper we shall apply, in this
specific context, the Field Strength Correlator Method [1], in the
framework of the SVM, as has been formulated in the preceding
paper ({\bf I}), {\it i.e.} in terms of its Worldline casting. In
particular, we shall study a simulated case of a meson-meson
scattering process whose quark-based description is of the general
form
\[
(1\bar{1})+(2\bar{2})\to(3\bar{3})+(4\bar{4})
\]
adopting a standard picture, already employed in the QCD
literature -see, for example [2] and [3], according to which quark
1 from the first meson and antiquark $\bar{2}$ from the second
meson are very heavy, in comparison to the incoming, total energy
$s$ -hence their worldlines are considered to remain intact from
the gluon field action. In turn this means that they can be
described in the framework of the eikonal approximation. The light
pairs $\bar{1},2$ and $\bar{3},4$, on the other hand, are
annihilated and produced in the $t$-channel, where the eikonal
approximation is not valid and a full treatment is called for
their description. In the Worldline framework the process is
schematically pictured in space-time by the straight eikonal lines
$(1\to 3)$ and $(\bar{2}\to\bar{4})$, describing an intact quark
and anti-quark and by the curves $(\bar{1}\to 2)$ and $(\bar{3}\to
4)$ which correspond, respectively, to the annihilated and
produced quark antiquark pairs. The structure of the  field
theoretical amplitude can be written as follows , see Fig.
\begin{equation}
G(x_4,x_3,x_2,x_1)=\langle iS_F(x_4,x_3\mid {\cal A}
iS_F(x_3,x_1\mid {\cal A})iS_F(x_1,x_2\mid {\cal
A})iS_F(x_2,x_4\mid {\cal A})\rangle_{\cal A}.
\end{equation}
In the above expression $iS_F$ is the full fermionic propagator
which, in the framework of the Worldline formalism, assumes the
form [4]
\begin{equation}
iS_F(y,x\mid {\cal A})=\int_0^\infty dL\, e^{-Lm^2}
\int\limits_{{\stackrel
{x(0)=x}{x(L)=y}}}Dx(\tau)e^{-{1\over4}\int\limits_0^Ld\tau\dot{x}^2}
\left[m-\frac{\gamma\cdot\dot{x}(L)}{2}\right]\Phi^{(1/2)}(L,0)P\exp\left(i\int\limits_0^Ld\tau\dot{x}\cdot
\cal{A}\right),
\end{equation}
where $\Phi^{(j)}$ is the so-called spin factor (see paper {\bf
I}) for the matter particles entering the system. For us, it means
that $j={1\over2}$.

Inserting the above formula into Eq. (1) we find
\begin{eqnarray}
&&G(x_4,x_3,x_2,x_1)=\prod\limits_{i=1}^4\int\limits_0^\infty
d\tau_i\theta(\tau_i-\tau_{i-1})e^{-(\tau_i-\tau_{i-1})m_i^2}
\int\limits_{{\stackrel {x(0)=x_4}{x(\tau_4)=x_4}}}\int
Dx(\tau)\delta[x(\tau_3)-x_3]\delta[x(\tau_2)-x_1]\times\nonumber\\
 &&\times\delta[x(\tau_1)-x_2]
\exp\left[-{1\over4}\int\limits_0^{\tau_4}
d\tau\,\dot{x}^2(\tau)\right]({\rm spin
\,\,structure})\left\langle P\exp\left(i\oint_C dx\cdot {\cal
A}\right)\right\rangle_{\cal A},
\end{eqnarray}
where the term $spin\,\, structure$ corresponds to the following
expression
\begin{equation}
({\rm spin\,\, structure})
=\prod\limits_{i=4}^1\left[m_i-{1\over2}\gamma\cdot{x}(\tau_i)\right]\Phi^{(1/2)}(\tau_i,\tau_{i-1}),\,\,(\tau_0\equiv
0).
\end{equation}

In principle, the Wilson loop appearing in Eq. (3) incorporates
the dynamics (perturbative, as well as non-perturbative) of the
process. In the framework of the Stochastic Vacuum Model (SVM) it
assumes the form (see {\bf I})

\begin{equation}
\left\langle P\exp\left(i\oint_C dx\cdot {\cal
A}\right)\right\rangle_{\cal A}=\exp\left[-{1\over
2}\int\limits_{S(C)}dS_{\mu\nu}(z)\int\limits_{S(C)}dS_{\lambda\rho}(z')\Delta^{(2)}_{\mu\nu,
\lambda\rho}(z-z')\right]\equiv e^{-A[C]}.
\end{equation}

The task to be undertaken in the present paper is to calculate the
amplitude (3), using the above expression which, it is reminded,
gives the structure of the Wilson loop in framework of the SVM.
The particular strategy to be adopted in our relevant effort can
be outlined as follows. In Section 2 we perform a semiclassical
calculation based on a combined minimization of the action $
A[C]$, see Eq. (17) of {\bf I}, with respect to the surface $S[C]$
and of the surface $S[C]$ with respect to the boundary $C$. This
procedure will allow us to determine the dominant contribution to
the Worldline integral (3) in the stochastic limit
$T_g^2\sqrt{\Delta}\ll 1$, as determined in {\bf I}.

As also noted in {\bf I}, the first order approximation of the
action ${\cal A}[C]$, in terms of the correlation length $T_g$, is
essentially the Nambu-Goto string action. The next order
corrections give rise to terms which  reveal different geometric
characteristics of the surface $S[C]$, given its embedding in a
4-dimensional background, such as, {\it e.g.}, its extrinsic
curvature. The presence of these terms, the origin of which is
completely different from the known quantum corrections of the
Nambu-Goto string, points out the powerful structure of the SVM.

In Section 2 we proceed further to take into account the rigidity
of the surface $S[C]$, which, as it will turn out, plays an
important role for  determining the reggeon intercept. Of lesser
importance, but in any case computable, are the corrections
related to the fluctuations of the  (Wilson) `curve' which forms
the boundary of the surface and will be discussed in Section 3. In
the same section we shall consider the contribution of the
spin-factor. As a point of note it should be mentioned that, in
order to compare our results with standard phenomenology, we shall
adopt some lattice-based parametrizations of the two-point
correlator. The technicalities of this matter will be discussed in
an Appendix.

\vspace*{1cm}

{\bf 2. Semiclassical Calculation}

According to Eq (3), in order  to obtain the full amplitude it
does not suffice to determine the minimal surface bounded by a,
given, specific contour -a problem which, in general, is very hard
to solve\footnote{Indeed, known cases for which solutions have
been found are limited and rather simple.}. One  needs to proceed
even further and sum over all possible boundaries with a weight of
the form
\begin{equation}
S[x]={1\over4}\int\limits_0^{\tau_4}d\tau\,\dot{x}^2+ A[C].
\end{equation}
The particular method we shall follow for getting an estimate of
the scattering amplitude is the minimization of the correlator
contribution to the action, {\it i.e.} $A[C]$, related to the
contour $C$, while at the same time find the minimal surface
corresponding to this specific boundary. In this way one obtains,
in principle, a result which enables one to determine the dominant
contribution to the path integral (3).

As was proved in {\bf I} the variation of $A[C]$ under changes of
the boundary reads:
\begin{equation}
g_\mu(\tau)=\frac{\delta A[C]}{\delta
x_\mu(\tau)}=\dot{x}_\alpha(\tau)\int\limits_{S(C)}
dS_{\gamma\delta}(z')\Delta^{(2)}_{\alpha\mu,\gamma\delta}[x(\tau)-z(\tau',s')]
\end{equation}
Accordingly, the correlator contributions become stationary for
the ``classical" trajectory
\begin{equation}
g_\mu[x_{{\rm cl}}]=0.
\end{equation}

Using the expansion for the correlator according to Eq(27) of the
previous paper, it is easy to see that
\begin{equation}
g_\mu(\tau)=2\dot{x}_\alpha(\tau)
R_{\alpha\mu}[x]-{1\over2}\dot{x}_\alpha(\tau)Q_{\alpha\mu}[x],
\end{equation}
with
\begin{equation}
R_{\alpha\mu}[x]=\int\limits_{S(C)}dS_{\alpha\mu}(z')D[x(\tau)-z(s',\tau')]
\end{equation}
and
\begin{equation}
Q_{\alpha\mu}[x]=\int
d\tau'\left[\dot{x}_\mu(\tau')\left(x_\alpha(\tau')-x_\alpha(\tau)\right)-(\mu\leftrightarrow
\alpha)\right]D_1[x(\tau)-x(\tau')],
\end{equation}
where the quantities $D$ and $D_1$ are introduced in the framework
of the SVM [1]. Their significance is of practical importance, as
far as the credibility of the SVM is concerned. Our eventual
numerical estimates in this work will use them as basic input. It
should be further noted that the above expressions are
reparametrization invariant. Also, in the last relation the
integration covers the whole range of the $\tau$ variable. Our
next step is to specify the minimal surface relevant to the
problem under study.

 Following Ref
[3] the minimal surface bounded by two infinite rods at a relative
angle $\theta$, has(in four-dimensional Euclidean space) the shape
of a (three-dimensional) helicoid, which is the only surface that
can be spanned by straight lines. In the considered process the
eikonal lines $1\to3,\,\bar{2}\to\bar{4}$, play the role of the
$`$rods', while the angle $\theta$ is connected, via analytic
continuation, to the logarithm of the incoming energy.

Given the above specifications, consider the following, helpful,
parametrization of the boundary $C$: For $0<\tau<\tau_1$ we have a
straight line segment, $x^{(1)}$, going from the point $x_4$ to
the point $x_2$. Introducing, moreover, for convenience the {\it
length} $2T=\mid x_4-x_2\mid$ and reparametrizing according to
$\tau\to\frac{2T}{\tau_1}\tau-T$, we write
\begin{equation}
x_\mu^{(1)}=(\tau,0,0,0), \,\,-T<\tau<T,
\end{equation}
with  $x_\mu^{(1)}(-T)=x_4,\, x_\mu^{(1)}(T)=x_2$.

The second eikonal line $x^{(3)}(\tau),\,\tau_2<\tau<\tau_3$, goes
from the point $x_1$ to the point $x_3$ at a relative angle
$\theta$ with respect to $x^{(1)}$, while a distance $b$ (impact
parameter) separates the two linear contours in a transverse
direction. Introducing the distance $2T_1=\mid x_3-x_1\mid$ and
reparametrizing according to
\begin{equation}
\tau\to
T_1\left(\frac{2}{\tau_3-\tau_2}\tau-\frac{\tau_3+\tau_2}{\tau_3-\tau_2}\right)
\end{equation}
we write
\[
x_\mu^{(3)}(\tau)=(-\tau\cos
\theta,\,-\tau\sin\theta,\,b,\,0),\quad -T_1<\tau<T_1,
\]
with $x_\mu^{(3)}(-T_1)=x_1,\,x_\mu^{(3)}(T_1)=x_3.$

In the following we shall assume, just for convenience, that
\[
2T=\mid x_4-x_2\,\mid\sim \,\mid x_3-x_1\mid=2T_1.
\]
For $\tau_1<\tau<\tau_2$, we have a helical curve
$x_\mu^{(2)}(\tau)$, which joins the points
$x_2=x_\mu^{(2)}(\tau_1)$ and $x_1=x_\mu^{(2)}(\tau_2)$,
representing the exchanged light quarks. Performing, now, the
change
\newline
$s=\frac{b}{\tau_2-\tau_1}(\tau-\tau_1)$, we write
\begin{equation}
x_\mu^{(2)}(s)=\left(\phi(s)\cos\frac{\theta s}{b},\,
\phi(s)\sin\frac{\theta s}{b},s,\,0\right),\,\,0<s<b.
\end{equation}

The continuity of the boundary requires
\[
x_\mu^{(1)}(T)=x_\mu^{(2)}(0)=x_2\quad {\rm
and}\,\,x_\mu^{(2)}(b)=x_\mu^{(3)}(-T)=x_1,
\]
or
\begin{equation}
\phi(0)=\phi(b)=T.
\end{equation}
The final helical curve is $x^{(4)}(\tau)$, which, for
$\tau_3<\tau<\tau_4$, joins the points $x_3=x^{(4)}(\tau_3)$ and
$x_4=x^{(4)}(\tau_4)$. Making one more, final, parametrization,
namely $s=\frac{b}{\tau_4-\tau_3}(\tau-\tau_3)$ we write
\begin{equation}
x_\mu^{(4)}(s)=\left(-\phi(s)\cos {\theta s\over b},\,-\phi(s)\sin
{\theta s\over b},\,s, 0\right),\,\,0<s<b.
\end{equation}
Once again, Eq. (15) takes care of the continuity of the boundary.
Now, the minimal surface is bounded by the (four) curves specified
by Eqs, (12)-(16) and can be spanned by straight lines
parametrized as follows
\begin{equation}
z_\mu(\xi)=\frac{T-\tau}{2T}x_\mu^{(4)}(s)+\frac{T+\tau}{2T}x_\mu^{(2)}(s)
=\left({\tau\over T}\phi(s)\cos{\theta s\over b},\,{\tau\over
T}\phi(s)\sin{\theta s\over b},\,s,0\right).
\end{equation}
It can be easily proved that the surface defined by the above
equation is minimal, irrespectively of the function $\phi$:
\begin{equation}
\partial_\tau\left[\frac{(\dot{z}\cdot
z')z'_\mu-z'^2\dot{z}_\mu}{\sqrt{g}}\right]+
\partial_s\left[\frac{(\dot{z}\cdot
z')\dot{z}_\mu-\dot{z}^2z'_\mu}{\sqrt{g}}\right]=0.
\end{equation}

One observes that the minimization of the surface is not enough
for the complete specification of the parametrization of the
helicoid. Accordingly, we go back to Eq. (8), which determines the
boundary that dominates the path integration (3). A first
observation is that, due to the antisymmetric nature of
$R_{\alpha\mu}$ and $Q_{\alpha\mu}$, the function $g_\mu$ vanishes
when $x_\mu(\tau)$ represents a straight line. Thus Eq. (8) is
trivially satisfied for the eikonal sector of the boundary.
Non-trivial contributions are coming only from the helices
$x_\mu^{(2)}$ and $x_\mu^{(4)}$. One can simplify Eq. (9) by
computing the leading behavior of the functions $R_{\alpha\mu}$
and $Q_{\alpha\mu}$ using the fact that the functions $D$ and
$D_1$, as defined in the SVM scheme -and measured in lattice
calculations [5]- decay exponentially fast for distances which are
large in comparison with the correlation length $T_g$ [1].
 In this connection and upon writing
 \[
 x(s')=x(s)+(s'-s)\dot{x}(s)+{1\over2}(s'-s)^2\ddot{x}(s)+\cdot\cdot\cdot,
 \]
 we find, for the second term in Eq. (9),
\begin{eqnarray}
\dot{x}_\alpha Q_{\alpha\mu}&=& {1\over
2}\left[(\dot{x}^2)\ddot{x}_\mu-(\dot{x}\cdot\ddot{x})\dot{x}_\mu\right]
\int\limits_0^b
ds'(s'-s)^2D_1\left[\dot{x}^2\frac{(s'-s)^2}{T^2_g}\right]
+\cdot\cdot\cdot=\nonumber\\
&=&{1\over \mid
\dot{x}\mid}\left(\ddot{x}-\frac{\dot{x}\cdot\ddot{x}}{\dot{x}^2}\right)\frac{1}{T_g\alpha_1}+\cdot\cdot\cdot,
\end{eqnarray}
where\footnote{We have omitted terms suppressed by powers of
$T_g^2$}
\[ \frac{1}{\alpha_1}\equiv T^4_g\int\limits_0^\infty
dw\,w^2D_1(w^2).
\]

 Noting that
\begin{eqnarray}
&& z_\mu(s,\tau=T)= x_\mu^{(2)}(s),\quad
z_\mu(s,\tau=-T)=x_\mu^{(4)}(s)\nonumber\\
&&\partial_\tau z_\mu(s,\tau)=\dot{z}_\mu(s,\tau)={1\over
2T}[x_\mu^{(2)}(s)-x_\mu^{(4)}(s)],
\end{eqnarray}
the leading behavior of the first term of the rhs of (9) can be
easily determined. One finds
\begin{eqnarray}
\dot{x}_\alpha
R_{\alpha\mu}&=&{1\over2}\dot{x}^2\left(\dot{z}_\mu-
\frac{(\dot{x}\cdot\dot{z})}{\dot{x}^2}{\dot{x}_\mu}\right)\int\limits_{-T}^T
d\tau'\int\limits_0^b ds'\,D\left[\dot{x}^2
\frac{(s'-s)^2}{T_g^2}\right]+\cdot\cdot\cdot\nonumber\\
&=&2T\mid\dot{x}\mid\left(\dot{z}_\mu-\frac{(\dot{x}\cdot\dot{z})}{\dot{x}^2}\dot{x}_\mu\right){\mu^2\over
T_g}+\cdot\cdot\cdot,
\end{eqnarray}
where we have introduced the parameter
\begin{equation}
\mu^2\equiv T^2_g\int\limits_0^\infty dw\,D(w^2).
\end{equation}
Thus, the function $g$ takes, to leading order, the form
\begin{equation}
g_\mu=\frac{1}{\mid\dot{x}\mid T_g}\left[4T\mu^2\dot{x}^2
\left(\dot{z}_\mu-\frac{\dot{x}\cdot\dot{z}}{\dot{x}^2}\dot{x}_\mu\right)
-\frac{1}{2\alpha_1}\left(\ddot{x}_\mu-\frac{\dot{x}\cdot\ddot{x}}{\dot{x}^2}\dot{x}_\mu\right)\right].
\end{equation}

Now, we recall from its definition, {\it cf} Eq. (7) that the
$g$-function provides a measure of the change of $A[C]$ when the
Wilson contour is altered as a result of some interaction which
reshapes its geometrical profile. In this sense, it contains
important information concerning the dynamics of the problem under
study. The structure of the $g$-function, as it appears in the
above equation, is quite general and exhibits its dependence, not
only on the boundary but on the minimal surface as well. It is
worth noting that this fact is strictly associated with the
non-Abelian nature of the theory since the function $D$ -and
consequently $\mu^2$- disappears [1] in an Abelian gauge theory.

Taking into account that for the helicoids parametrization the
velocity $\dot{x}$ has three non-zero components, while $\ddot{x}$
and $\dot{z}$ have only two, we conclude that Eq. (8) can be
satisfied only if
\begin{equation}
4T\mu^2\dot{x}^2\dot{z}_\mu-{1\over 2\alpha_1}\ddot{x}_\mu=0.
\end{equation}
Inserting in Eq. (24) the helical parametrization one easily finds
that the function $\phi$ must be a constant. Taking, now, into
account Eq. (14) we determine this constant to be the length $T$.
It is then very easy to see that this result leads to the
conclusions
\begin{equation}
\dot{x}\cdot\dot{z}=0,\,\,\dot{x}\cdot\ddot{x}=0
\end{equation}
and
\begin{equation}
\dot{x}^2=-\frac{1}{8\mu^2\alpha_1}\frac{\theta^2}{b^2},
\end{equation}
where
\[
\dot{x}^2=1+\frac{T^2\theta^2}{b^2}.
\]

 This equation cannot
be satisfied in Euclidean space. In Minkowski space the angle
$\theta$ becomes imaginary $\theta\to-i\chi\simeq\ln\left({s\over
2m^2}\right)$
 and Eq. (26) has
a positive definite solution:
\begin{equation}
\frac{T^2\chi^2}{b^2}=1-\frac{1}{8\mu^2\alpha_1}\frac{\chi^2}{b^2}.
\end{equation}
The above formula relates the impact parameter $b$, the logarithm
of the incoming energy $\ln\left(\frac{s}{2m^2}\right)$ and the
distance $T$. These parameters must not be considered as
independent from each other in a calculation of the leading
behavior of the scattering amplitude. In fact, Eq. (27) indicates
that the effective impact parameter must grow with the incoming
energy: $b\sim\ln s$, a conclusion which is in agreement with the
landmark result of Cheng and Wu [6].

The preceding analysis, obviously repeats itself for the two
helical curves
 $x^{(2)}$ and $x^{(4)}$ and has led us to a specific parametrization
 for the Wilson loop, which plays the dominant role in the path
 integration in Eq. (3). We are now in position to determine the leading contribution to
 the action (6):
 %\end{document}
\begin{equation}
S_{\rm cl}={1\over4} \int\limits ^{\tau_4}_0
d\tau\dot{x}^2_{cl}(\tau)+A[C]_{\rm cl}.
\end{equation}
Our first step is to expand the second term of the integrand in
powers of $T^2_g\sqrt{\Delta}$. The first term of such an
 expansion is the familiar Nambu-Goto string. The next term,
 which reveals the rich structure of the SVM, is the so-called
 `rigidity term', representing the extrinsic curvature of a
 surface embedded  in a four-dimensional [7] background:
\begin{equation}
A[C]=\sigma\int d^2\xi\sqrt{g}+{1\over\alpha_0}\int
d^2z\xi\sqrt{g}g^{ab}\partial_a t_{\mu\nu}\partial_b
t_{\mu\nu}+\cdot\cdot\cdot,
\end{equation}
where, in the above expression,
\begin{equation}
\sigma\equiv{1\over2}T^2_g\int d^2zD(z^2)
\end{equation}
enters as the string tension.

The coefficient of the rigidity term reads
\begin{equation}
{1\over\alpha_0}\equiv{1\over32}T^4_g\int
d^2z\,z^2(2D_1(z^2)-D(z^2)).
\end{equation}
Terms proportional to $T_g^6$ entering the expansion in Eq. (28)
will be considered negligible in our analysis. We have also
omitted the term $\int d^2\xi\sqrt{g}R$, since in two dimensions
the curvature is a total derivative. Using the helicoids
parametrization (17), with $\phi=T$, the Nambu-Goto term in Eq.
(29) takes the form
\begin{equation}
\int d^2\xi\sqrt{g}=\int\limits_{-T}^Td\tau\int\limits_0^b
ds\sqrt{1+\frac{\tau^2\theta^2}{b^2}}
=bT\left[\sqrt{1+p^2}+{1\over
p}\ln\left(\sqrt{1+p^2}+p\right)\right],
\end{equation}
where $p=\frac{T\theta}{b}$.

To proceed further we analytically continue to Minkowski space
where we can use Eq. (27) to determine
\begin{equation}
bT\sqrt{1+p^2}\to bT\sqrt{1-\frac{T^2\chi^2}{b^2}}\simeq
b\left(1-\frac{1}{8\alpha_1\mu^2}\frac{\chi^2}{b^2}\right)^{1/2}\frac{1}{
\sqrt{8\alpha_1\mu^2}}\simeq \frac{b}{
\sqrt{8\alpha_1\mu^2}}+{\cal O}(T^3_g)
\end{equation}
and
\begin{equation}
\frac{bT}{p}\ln\left(\sqrt{1+p^2}+p\right)\to
\frac{bT}{-iT\chi/b}\ln\left[\sqrt{1-\frac{T^2\chi^2}{b^2}}
-i{T\chi\over b}\right]\simeq \frac{\pi
b^2}{2\chi}-\frac{b}{\sqrt{8\alpha_1\mu^2}}+ {\cal O}(T^3_g).
\end{equation}
Thus
\begin{equation}
\sigma\int d^2\xi\sqrt{g}\to\frac{\sigma\pi b^2}{2\chi}.
\end{equation}
In the same framework, the contribution of the rigidity term takes
the form
\begin{eqnarray}
\int d^2\xi\sqrt{g}\,g^{ab}\partial_a\,
t_{\mu\nu}\partial_b\,t_{\mu\nu} &=& \int\limits_{-T}^T
d\tau\int\limits_0^b ds
\frac{1}{\sqrt{1+\frac{\theta^2\tau^2}{b^2}}}\left(\frac{\theta^2}{b^2}+{1\over2}
\frac{\theta^4}{b^4}\tau^2\right)\nonumber\\&=&\theta
\left[{3\over2}\ln\left(\sqrt{1+p^2}+p\right)+{1\over2}p
\sqrt{1+p^2}\right].
\end{eqnarray}

It follows that in Minkowski space we have
\begin{equation}
{1\over\alpha_0}\int
d^2\xi\sqrt{g}g^{ab}\partial_at_{\mu\nu}\partial_ bt_{\mu\nu}\to
-\frac{3\pi}{4\alpha_0}\chi.
\end{equation}
For the full estimation of the classical action, {\it cf} Eq.
(28), one should also take into account the presence of the
classical kinetic term. Non trivial contributions come from the
helical curves $x^{(2)}(\bar{1}\to 2)$ and $x^{(4)}(\bar{3}\to
4)$:
\begin{equation}
\frac{b}{4(\tau_2-\tau_1)}\int\limits_0^b ds(\dot{x}^{(2)})^2+
\frac{b}{4(\tau_4-\tau_3)}\int\limits_0^b ds(\dot{x}^{(4)})^2
=\frac{b^2\dot{x}^2}{4(\tau_2-\tau_1)}+\frac{b^2\dot{x}^2}{4(\tau_4-\tau_3)}.
\end{equation}
Now we have to take into account that both $\tau_2-\tau_1$ and
$\tau_4-\tau_3$ must be integrated with weights
$e^{-(\tau_2-\tau_1)m^2}$ and $e^{-(\tau_4-\tau_3)m^2}$,
respectively. These integrals, as it turns out, are dominated by
the values
$\tau_2-\tau_1=\tau_4-\tau_3=\frac{b\mid\dot{x}\mid}{2m}$, leading
to a final kinetic contribution of the form
\begin{equation}
2mb\mid\dot{x}\mid=2\frac{m}{\sqrt{8\alpha_1\mu^2}}\chi.
\end{equation}
Here, $m$ is the mass of the light quarks, thus the result
expressed by (34) can be considered negligible.

 From the above analysis we conclude that
\begin{equation}
S_{\rm cl}\approx\frac{\sigma\pi
b^2}{2\ln\left({s\over2m^2}\right)}-\frac{3\pi}{4\alpha_0}\ln\left(\frac{s}{2m^2}\right).
\end{equation}
Putting aside, for now, the possible corrections to $A[C]$ which
arise from fluctuations of the boundary as well as the spin factor
contribution, let us consider the result (39) as a whole, except
for terms $\sim  m$. To obtain the final expression for the
scattering amplitude one must integrate over the impact parameter:
\begin{equation}
\int
d^2b\exp\left(i\vec{q}\cdot\vec{b}-\frac{\sigma\pi}{2\chi}b^2\right)\propto\exp\left(-{1\over2\pi\sigma}q^2\chi\right).
\end{equation}

Combining (40) and (41) we find, for the scattering amplitude, a
Regge behavior of the form $s^{\alpha'_R(0)t+\alpha_R(0)}$ with
\begin{equation}
\alpha'_R(0)={1\over2\pi\sigma}\,\,{\rm
and}\,\,\alpha_R(0)=\frac{3\pi}{4\alpha_0}.
\end{equation}

In the Appendix we present a certain parametrization [8] for the
functions $D$ and $D_1$ entering the SVM scheme which give for the
string tension the value $\sigma \approx 0.175$ GeV$^2$ and for
the coefficient of the rigidity term the value
$\frac{1}{\alpha_0}\approx 0.276$. With these numbers we obtain
for the Regge slope the value $\alpha'_R(0)\approx 0.91$
GeV$^{-2}$ and for the Reggeon intercept the value
$\alpha_R(0)\approx0.65$, in good agreement with the
phenomenological values $\alpha'_R(0)= 0.93$ GeV and
$\alpha_R=0.55$. [3]
\vspace*{.3cm}

{\bf 3. Boundary Fluctuations and the Role of the Spin Factor}
\vspace*{.3cm}

As repeatedly mentioned in our narration, corrections to the
amplitude (3), beyond semiclassical ones, are expected to arise
from fluctuations of the boundary of the surface on which the
two-point correlator `lives'. Fluctuations of the surface itself
can be taken into account by higher order correlators. This, in
fact, is the big difference which distinguishes the SVM approach,
in comparison with Nambu-Goto type approaches.

We begin our related considerations by expanding the action (6)
around the helicoid classical solution:
\begin{eqnarray}
&&S=S_{{\rm
cl}}-{1\over2}\int\limits_0^{\tau_4}d\tau\,y(\tau)\ddot{x}^{{\rm
cl}}(\tau)+{1\over2}\int\limits_0^{\tau_4}d\tau\int\limits_0^{\tau_4}d\tilde{\tau}\,
y_\alpha\tau)\times\nonumber\\&&
\times\left[-{1\over2}\delta_{\alpha\beta}\frac{\partial^2}{\partial\tau^2}\delta(\tau-\tilde{\tau})
+\frac{\delta^2A[C]}{\delta x^{{{\rm cl}}}_\alpha(\tau)\delta
x_\beta(\tilde{\tau})}\right]y_\beta(\tilde{\tau})+\cdot\cdot\cdot,
\end{eqnarray}
where $y=x-x^{\rm cl}$.

 Using the results of {\bf I} one can easily determine that
\begin{eqnarray}
\frac{\delta^2A[C]}{\delta x_\alpha(\tau)\delta
x_\beta(\tilde{\tau})}=&&\dot{x}_\mu(\tau)\dot{x}_\nu(\tilde{\tau})
\Delta^{(2)}_{\mu\alpha,\nu\beta}[x(\tau)-x(\tilde{\tau})]-\nonumber\\&&
-\frac{\partial}{\partial\tau}\delta(\tau-\tilde{\tau})\int\limits_{S(C)}dS_{\lambda\rho}(z')
\Delta^{(2)}_{\alpha\beta,\lambda\rho}[z(\xi'-x(\tau)]+\nonumber\\&&
+\dot{x}_\alpha(\tau)\int
ds\,\alpha(\tilde{\tau},s)\dot{z}_\lambda(\tilde{\tau},s)z'_\rho(\tilde{\tau},s)\epsilon^{\kappa\nu\lambda\rho}
\Delta_{\kappa\alpha\mu}[z(\tilde{\tau},s)-x(\tau)],
\end{eqnarray}
where we have written
\[
\frac{\delta z_\mu(\tau,s)}{\delta
x_\nu}=\delta_{\mu\nu}\delta(\tau-\tilde{\tau})a(\tilde{\tau},s).
\]
The second term on the rhs of Eq.(44) is simply the area
derivative which, as we have seen in {\bf I}, has the general form
$\frac{\delta A[C]}{\delta\sigma_{\alpha\beta}}\sim
g_{\alpha}\dot{x}_\beta-g_\beta\dot{x}_\alpha$. Thus, for the
classical solution $g[x^{\rm cl}]$ it gives zero contribution. It
is, furthermore, easy to verify that the third term in (44) also
disappears for $x=x^{\rm cl}$. We, therefore, conclude that
\begin{equation}
\frac{\delta^2A[C]}{\delta x_\alpha^{\rm cl}(\tau)\delta
x_\beta^{\rm cl}(\tilde{\tau})}=\dot{x}_\mu^{\rm
cl}(\tau)\dot{x}_\nu^{\rm
cl}(\tilde{\tau})\Delta^{(2)}_{\mu\alpha\nu\beta}[ x^{\rm
cl}(\tau)-x(\tilde{\tau})].
\end{equation}

Inserting Eq. (45) into Eq. (44) and taking into account that the
dominant contribution to the two-point correlator comes from the
region $\tau\approx\tilde{\tau}$ we find
\begin{equation}
S\approx S_{\rm{cl}}+\int\limits_0^b
ds\,y_\alpha(s)\left[-{1\over2}\frac{m}{\mid\dot{x}\mid}\delta_{\alpha\beta}\frac{\partial^2}{\partial
s^2}+\frac{\lambda^2}{T_g}\omega_{\alpha\beta}(s)\right]y_\beta(s).
\end{equation}
Let it be remarked that to arrive at the above relation we have
adopted the expansion of the two-point correlator indicated in Eq.
(21) of paper {\bf I}. We have also used the helicoid
parametrization observing, at the same time, that the eikonal
lines give null contribution. One further realizes that the
contributions of the two helical curves to the linear term in (42)
cancel each other, since
$\ddot{x}^{(2)}_\mu(s)=-\ddot{x}^{(4)}_\mu(s)$ and $\tau_2-\tau_1
\simeq \tau_4-\tau_3\sim\frac{b\mid\dot{x}\mid}{2m}$.

The non-trivial contribution of the helical curves is incorporated
in the term
\begin{equation}
\omega_{\alpha\beta}=\delta_{\alpha\beta}-{1\over2\dot{x}^2}
\left(\dot{x}_\alpha^{(2)}\dot{x}_\beta^{(2)}+\dot{x}_\alpha^{(4)}\dot{x}_\beta^{(4)}\right),
\end{equation}
the origin of which is the second functional derivative, {\it
c.f.} (48). The mass parameter $\lambda^2$ in (46) has the same
source and is defined as
\begin{equation}
\lambda^2\equiv \mid \dot{x}\mid T_g^2\int\limits _0^\infty
dw\left(D(w^2)+D_1(w^2)+{d\over dw^2}D_1(w^2)\right).
\end{equation}

The differential operator entering Eq. (46) has no zero
eigenvalues since the ``classical'' solution is, in fact, the one
that annihilates the $g$-function. Accordingly, the calculation of
the path integral over $y=x-x^{{\rm cl}}$ does not require any
particular regularization. A straightforward calculation shows
that
\begin{equation}
\det\omega_{\alpha\beta}={1\over\dot{x}^2}\left(1-{1\over\dot{x}^2}\right)
=\frac{T^2\theta^2/b^2}{1+\theta^2/b^2}.
\end{equation}
Thus the matrix $\omega_{\alpha\beta}$ can be diagonalized and the
$y$-integral can be easily performed. However, in the limit $m\to
0$ it can be immediately seen that the integration over the
boundary fluctuations gives prefactors which are powers of the
logarithm of the incoming energy and as far as Regge behavior is
concerned, they cannot change the behavior that was determined in
the previous section.

The next task is to take up the issue of the spin-field dynamics
contribution to the scattering amplitude. As seen in {\bf I}, a
spin factor is associated with each segment of the worldline path.
This factor receives contributions from two sources. The first one
is
\begin{equation}
\int d\tau\int\limits_{S(C)} dS\cdot\Delta^{(2)}(z-x)\cdot J=\int
d\tau\frac{\dot{x}_\mu g_\nu-\dot{x}_\nu
g_\mu}{\dot{x}^2}{i\over4}\left[\gamma_\mu,\gamma_\nu\right]
\end{equation}
and is obviously zero for the classical trajectory (8)

The other term has the form
\begin{equation}
S={1\over8}\int d\tau\int
d\tau'J_{\mu\nu}\Delta^{(2)}_{\mu\nu,\lambda\rho}(x-x')J_{\lambda\rho}={3\over4}\int
d\tau\int d\tau'(D+D_1)+{3\over8}\int d\tau\int
d\tau'(x-x')^2D'_1.
\end{equation}

In the stochastic limit, within which we are working, the
integrals in the above equation give appreciable contribution to
(51) only for $\mid
x(\tau)-x(\tau')\mid\approx\mid\dot{x}\mid\,\mid\tau-\tau'\mid\ll
T_g$. More concretely, consider the contribution to (51) from the
helical curve $(\bar{1}\to 2)$. A straightforward calculation
shows that the analytically continued result is
\begin{equation}
S=-(t_2-t_1)^2{M^4\over\chi},
\end{equation}
where we have written $\tau=it$ for the time variable and denoted
\begin{equation}
M^4=\left(8\frac{\int\limits_0^\infty dwD(w)}{\int\limits_0^\infty
dw\,w^2D_1(w)}\right)^{1/2} \int\limits_0^\infty
dw\left(D(w^2)+D_1(w^2)+{1\over2}\frac{d}{dw^2}D_1(w^2)\right).
\end{equation}

As has been mentioned in {\bf I} and discussed in [9],
contribution (52) has an interesting role as far as the form of
the fermionic propagator is concerned, but it is obvious that it
does not alter the basic Regge structure of the amplitude was
calculated in the previous section.

The remaining spin structure is summarized in the chain
\begin{equation}
I=\prod\limits_{i=4}^1m_i\left[1-{1\over2m_i}\gamma\cdot\dot{x}^{(i)}(\tau_i)\right],
\end{equation}
which must be sandwiched between the external spinor wavefunctions
representing the incoming and outgoing quarks (in the simple
picture wherein the meson wavefunction is just the product of free
spinors). The non-trivial dynamics of the process are now
incorporated into the fact that the vectors
$x_\mu^{(i)},\,i=1,2,3,4$ forming the boundary of the helicoids,
are 3-dimensional vectors with $\mid\dot{x}^{(i)}\mid^2=const.$
For $i=1,3$ turns the factor in (54) to the operator
$1-\frac{\gamma\cdot p^{(i)}}{\mid p^{(i)}\mid}$.

For $i=2,4$ the matrices
\begin{equation}
I_2=1-{1\over2m}\frac{b}{\tau_2-\tau_1}\gamma\cdot\dot{x}^{(2)}(b)\to1-\frac{\gamma\cdot\dot{x}^{(2)}(b)}{\mid
\dot{x}^{(2)}\mid}
\end{equation}
and
\begin{equation}
I_4=1-{1\over2m}\frac{b}{\tau_4-\tau_3}\gamma\cdot\dot{x}^{(4)}(b)\to1-\frac{\gamma\cdot\dot{x}^{(4)}(b)}{\mid
\dot{x}^{(4)}\mid},
\end{equation}
are also representations of projection operators. As shown in [3]
the matrices (55) and (56) are the direct product of two $2\times
2$ matrices each of which are by themselves projection operators.
Given these observations it becomes a matter of simple algebra to
find that the standard kinematics are reproduced.

\newpage

\vspace*{2cm}

\appendix
\setcounter{section}{0} \addtocounter{section}{1}
\section*{Appendix}
\setcounter{equation}{0}
\renewcommand{\theequation}{\thesection.\arabic{equation}}

\appendix
\setcounter{section}{0} \addtocounter{section}{1}
\section*{Appendix}
\setcounter{equation}{0}
\renewcommand{\theequation}{\thesection.\arabic{equation}}

 In this Appendix we  present a parametrization of the functions
 $D$ and $D_1$, already referred to in {\bf I} and used
 extensively in the present paper. This parametrization
  is supported by lattice data and is extensively discussed
 in Ref. [8].

 The exact relations defining the functions are
 \begin{equation}
  D=\frac{\pi^2(N^2_C-1)}{2N_C}\frac{G_2}{24}\kappa D_N,\quad
  D_1=\frac{\pi^2(N^2_C-1)}{2N_C}\frac{G_2}{24}(1-\kappa)D_{1,N},
 \end{equation}
 where $D_N$ and $D_{1,N}$ are functions which determine the
 structure of the two-point correlators, as defined in [8].
 The factor $G_2$ is defined as follows
 \begin{equation}
 G_2\equiv
 \langle0\mid\frac{g^2}{4\pi}F_{\mu\nu}^\alpha(0)F_{\mu\nu}^\alpha(0)\mid0\rangle
 =\frac{2N_C}{4\pi^4}\Delta^{(2)}_{\mu\nu,\mu\nu}(0).
\end{equation}
For the above correlator we shall adopt the value given in Ref
[8], namely $G_2=(0.496)^4GeV^4$. The value of the numerical
quantity $\kappa$ in (A.1) is estimated in the same reference to
be 0.74 .

The ansatz for the function $D_N$ is [8]
\begin{equation}
D_N(z)={27\over64}\,{1\over a^2}\int d^4k e^{ik\cdot
z}\frac{k^2}{\left[k^2+\left(\frac{3\pi}{8a}\right)^2\right]^4},
\end{equation}
where
\begin{equation}
a\equiv\int\limits_0^\infty dz\,D_N(z).
\end{equation}
 A simple calculation shows
that
\begin{equation}
D_N(z)=wK_1(w)-{1\over4}w^2K_0(w), \quad w=\frac{3\pi}{8a} \mid
z\mid,
\end{equation}
with $K_\nu$ denoting a Bessel function.
%\end{document}
The correlation length $T_g$, introduced in {\bf I} -and
frequently used in the text- can be deduced from Eq. (A.5):
\begin{equation}
T_g=\frac{8a}{3\pi}.
\end{equation}
The estimated value of $a$ is
\begin{equation}
a\approx 0.35\,{\rm fm}\,{\rm or} \,\,T_g\approx 0.297 \,{\rm fm}.
\end{equation}

With the help of ansatz (A.3) and using (A.7) one can determine
the string tension:
\begin{equation}
\sigma={1\over2}T^2_g\int
d^2wD(w)={1\over2}T^2_g\frac{\pi^2(N_C^2-1)}{2N_C}{G_2\over24}\kappa\int
d^2w\left[wK_1(w)-{1\over4}w^2K_0(w)\right],
\end{equation}
or
\begin{equation}
\sigma={1\over2}T^2_g
\frac{\pi^2(2N_C^2-1)}{2N_C}\frac{G_2}{24}\kappa2\pi=a^2G_2\kappa\pi{32\over81}\approx
0.175 GeV^2
\end{equation}

The anzatz for the function $D_{1,N}$is deduced from the equation
[8]
\begin{equation}
\left(4+z_\mu\frac{\partial}{\partial
z_\mu}\right)D_{1,N}(z)=4D_N(z)
\end{equation}
or
\begin{equation}
D_{1,N}(z)={1\over z^4}\int_0^z dw[4w^4K_1(w)-w^5K_0(w)]
\end{equation}
The coefficient of the rigidity term entering Eq. (31) can now be
calculated:
\begin{eqnarray}
{1\over\alpha_0}&=&{1\over32}T^4_g\int
d^2w\,w^2[2D_1(w)-D(w)]=\nonumber\\&&
={1\over32}T^4_g\frac{\pi^2(N_C^2-1)}{2N_C}{G_2\over24}\int
d^2w\,w^2[2(1-\kappa)D_{1,N}(w)-\kappa D_N(w)]\nonumber\\&&
={1\over32}T^4_g\frac{\pi^2(N_C^2-1)}{2N_C}\frac{G_2}{24}2(1-\kappa)32\pi\approx
0.276.
\end{eqnarray}

\begin{center}
{\bf Acknowledgement}
\end{center}

\vspace{0.2cm}

The authors wish to acknowledge financial supports through the
research program ``Pythagoras'' (grant 016) and by the General
Secretariat of Research and Technology of the University of
Athens.

We also acknowledge the assistance of our student N. Kaplis for
the drafting the figure appearing in the paper.

%%%%%%%%%%%%%%%%%%%%%%%%%%%%%%%%%%%%% BIBLIOGRAFIA %%%%%%%%%%%%%%%%%%%%%%%%%%%%%%%%%%%%%%%%%%%%%%

%%%%%%%%%%%%%%%%%%%%%%%%%%%%%%%%%%%%% BIBLIOGRAFIA %%%%%%%%%%%%%%%%%%%%%%%%%%%%%%%%%%%%%%%%%%%%%%


\begin{thebibliography}{99}

\bibitem{DO-SI 1}
H. G.Dosch, Phys. Lett. {\bf B}190(1987) 177; H. G.Dosch and Yu.
A. Simonov, Phys. Lett. {\bf B}205(1988)339; Yu. A. Simonov, Nucl.
Phys. {\bf B}307(1988)512; Yu. A. Simonov, Phys. Usp. 39
(1996)313; Usp. Fiz. Nauk. (1996) 337, hep-ph/9709344; A.
DiGiacomo, H. G. Dosch, V. A. Shevchenko and Y. Simonov, Phys.
Rep. {\bf372} (2002) 319, hep-ph/0007224.

\bibitem{J. Botts- Sterman 2} J. Botts and G Sterman Nucl. Phys.
{\bf B} 325 (1998) 62.

\bibitem{J-P 3} R. A. Janic and R. Peschanski  Nucl. Phys. {\bf
B}625(2002)279.

\bibitem{Karanikas-Ktorides 4} A. I. Karanikas and C. N. Ktorides,
Phys. Rev {\bf D 52} (1995) 5883.

\bibitem{DiGiac 5}A. DiGiacomo and H. Panagopoulos, Phys. Lett.
{\bf B} 285(1992)133; Del Debbio, A. DiGiacomo and Yu. A. Simonov,
Phys. Lett. {\bf B} 332(1994)111; A. DiGiacomo, M. Maggiore and S.
Olejnik Phys. Lett. {\bf B} 236(1990)199; Nucl. Phys. {\bf B} 347
(1990)441;A. DiGiacomo, M. D'Elia, E. Meggiolaro and H.
Panagopoulos, hep-lat/9808056.

\bibitem{Cheng-Wu 6} H. Cheng and T.T. Wu {\it Expanding
Protons}
MIT Press, Cambridge Mass. (1987) and original references cited
therein.

\bibitem {K-K 7} A. I. Karanikas and C. N. Ktorides MIT-CPT-1604,
June, 1988; D. V. Antonov, Yu. A. Simonov and D Ebert Surveys High
Energy Phys. 10, (1997) 421.

\bibitem{Nacht 8}
O. Nachtmann, {\it High Energy Collisions
and Non-Perurbative QCD}
in ``Perturbative and Non-Perurbative Aspects of Quantum Field
Theory'', H. Latal, W. Schwinger (Eds.) Springer-Verlag, Berlin,
Heidelberg (1997) and references therein.

\bibitem{Chet 9}K. G. Chetyrkin, S. Narison and V. I. Zakharov,
Nucl. Phys. {\bf B} 550(1999) 353; A. I. Karanikas and C. N.
Ktorides Phys. Atom. Nucl. 68(2005); Yad. Fiz. 68(2005)894,
hep-ph/0405077.

\end{thebibliography}
\end{document}